\documentclass[article]{elsarticle}

\usepackage{lineno,hyperref}
\usepackage{color}
\modulolinenumbers[50]

\journal{Journal of \LaTeX\ Templates}

\newcommand{\mcc}[1]{\multicolumn{1}{c}{#1}}    

\begin{document}

\begin{frontmatter}

\title{Antiferromagnetic majority voter model on square and honeycomb lattices}

\author[ifug]{Francisco Sastre}
\ead{sastre@ugto.mx}

\author[nancy]{Malte Henkel}
\ead{malte.henkel@univ-lorraine.fr}

\address[ifug]{
Departamento de Ingenier\'a F\'isica, Universidad de Guanajuato,
AP E-143, CP 37150, Le\'on, M\'exico
}

\address[nancy]{
Groupe de Physique Statistique, Institut Jean Lamour (CNRS UMR 7198), Universit\'e de Lorraine Nancy,
BP 70239, F -- 54506 Vand{\oe}uvre-l\`es-Nancy, France,
}

\begin{abstract}
An antiferromagnetic version of the well-known majority voter model on square and honeycomb lattices is proposed. 
Monte Carlo simulations give evidence for a continuous order-disorder phase transition in the stationary state in both cases. 
Precise estimates of the critical point are found from the combination of three cumulants, and our results are in
good agreement with the reported values of the equivalent ferromagnetic systems. The critical exponents $1/\nu$, $\gamma/\nu$ and
$\beta/\nu$ were found. Their values indicate that the stationary state of the antiferromagnetic majority voter model belongs to the Ising model
universality class.
\end{abstract}

\begin{keyword}
Antiferromagnetic,
Majority voter model,
Finite-size scaling,
Ising model, 
universality class
\end{keyword}

\end{frontmatter}

\linenumbers

\section{Introduction}

The Majority voter model (MVM) is a simple non-equilibrium Ising-like system, proposed as a way to simulate opinion dynamics. 
The collective behaviour of the voters shares many aspects with the well-established theory of non-equilibrium phase transitions and
results form simulations can be analysed similarly~\cite{Henkel08}. 
In the standard MVM, the system evolves following a dynamics
where each ``voter" (spin) assumes the same opinion as the majority of its neighbours, with probability $(1+x)/2$ 
and assumes the opposite opinion, with probability $(1-x)/2$. Here $x$ is
the control parameter, with a range $0\le x\le 1$. 
The stationary state of the MVM presents a second-order phase transition, at some critical value $x_c$, 
and previous numerical works on regular lattices show that the
critical exponents are compatible with its equivalent lattices for the Ising model~\cite{Oliveira1992,Kwak2007,AcunaLara2012,AcunaLara2014}. 
Those results seem to confirm the
conjecture that non-equilibrium models with up-down symmetry and spin-flip dynamics 
fall in the universality class of the Ising model~\cite{Grinstein1985}.
However, numerical simulations in non-regular latices, such Archimedean, small-world or random lattices, 
rather seem to indicate that the critical exponents are governed by the lattice
topology~\cite{Campos2003,Lima2005,Pereira2005,Lima2006,Luz2007,Lima2008,Santos2011,Crokidakis2012,Stone2015}.

On the other hand, the MVM belongs to a family of generalized spin models~\cite{Oliveira1993} that can be modelled 
in terms of a competing dynamics, induced by heat baths at two different
temperatures (on two dimensional square lattices)~\cite{Tamayo1994,Drouffe1999} and hence have a {\em non}-equilibrium stationary state. 
The equilibrium stationary state of the ferromagnetic Ising model is recovered when both temperatures are equal; 
in this case, the detailed-balance condition is fulfilled.
All members of the family share the same definition of the order-parameter and there is a critical line that 
separates the paramagnetic (disordered) phase from the ferromagnetic (ordered)
phase~\cite{Oliveira1992}. So far, the ferromagnetic-paramagnetic transition out of equilibrium has been extensively studied
with a single-flip dynamic rule that recovers the equilibrium Ising model. 
When antiferromagnetic interactions are included in the Ising model, the phenomenology becomes
richer, additional phases with different critical behaviour, multicritical points, etc. can appear 
(see Ref.~\cite{Binder1980} and references therein). In this work, we want
to explore, as a starting point,
if it is possible to implement an antiferromagnetic non-equilibrium version that follows a single-flip updating scheme. 
Godoy and Figureido~\cite{Godoy2002} proposed a non-equilibrium mixed-spin antiferromagnetic model with two updating
schemes: one single-spin via Glauber dynamics and a two-spin updating linked to an external energy source. 
The aim of the present work is to implement the antiferromagnetic version of the MVM on the square and the honeycomb two-dimensional lattices, 
evaluate the critical point and the stationary critical exponents, $\beta$, $\gamma$ and $\nu$.

This work is organised as follows: in section~2, the antiferromagnetic MVM is defined and the finite-size scaling method used to analyse its
stationary state is outlined. In section~3, the results of the Monte Carlo simulation are reported and the critical parameters are extracted. 
We conclude in section~4. 

\section{Model and Finite-Size Scaling}
In the ferromagnetic version of the MVM \cite{Oliveira1992}, 
each lattice site is occupied by a spin, $\sigma_i$, that interacts with its nearest neighbours. The system evolves in
the following way: during an elementary time step, a spin $\sigma_i = \pm 1$ on the lattice is randomly selected, and 
flipped with a probability given by
    \begin{equation}  \label{transition}
        p(x)=\frac{1}{2}(1+\eta x).
    \end{equation}
Herein, $\eta$ stands for the (ferromagnetic) rule
    \begin{equation} \label{2}
    \eta = \left\{ \begin{array}{ll} -\mbox{sgn}[H_{i}\cdot\sigma_{i}] & \mbox{\rm ~~;~ if $H_i\ne 0$} \\
                                     0                                 & \mbox{\rm ~~;~ if $H_i= 0$} 
                   \end{array} \right.
    \end{equation}
and where $H_i$ is the local field produced by the nearest neighbours to the $i$~th spin.
The control parameter $x$ acts analogously to a noise in the system. With the dynamics (\ref{transition},\ref{2}), 
a given spin $\sigma_i$ adopts the sign of $H_i$ (the majority of its nearest neighbours) with probability $(1+x)/2$ and the opposite sign of
$H_i$ (the minority) with probability $(1-x)/2$. 
In the Ising ferromagnet, the sign of the bilinear exchange interaction in the hamiltonian defines if the system is ferromagnetic,
or antiferromagnetic, respectively, for a positive or negative sign. 

An {\em anti}ferromagnetic version of the MVM is now naturally obtained by replacing the rule (\ref{2}) by 
    \begin{equation} \label{3}
    \eta = \left\{ \begin{array}{ll} +\mbox{sgn}[H_{i}\cdot\sigma_{i}] & \mbox{\rm ~~;~ if $H_i\ne 0$} \\
                                     0                                 & \mbox{\rm ~~;~ if $H_i= 0$} 
                   \end{array} \right.
    \end{equation}
in combination with the flip probability (\ref{transition}), with $0<x<1$. 
The anti-ferro MVM is defined by the dynamics (\ref{transition},\ref{3}), with the control parameter $0<x<1$. 
Alternatively, one could keep the rule (\ref{2}), but replace $x\mapsto -x$ in the flip probability (\ref{transition}). 
The consideration of square and honeycomb lattices allows to study the effect of having an even or odd number of nearest neighbours of each
spin on the critical behaviour in the stationary state. 

\begin{figure}[h!]
    \begin{center}
     \includegraphics[width=11.0cm,clip]{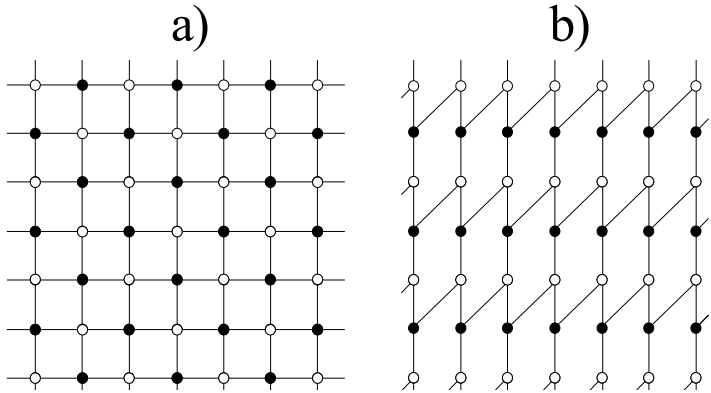}
     \caption{\label{redes}
     Sub-lattices for a) the square and b) the honeycomb geometries. Here, $c_i=1$ for white sites and $c_i=-1$ for black sites.
     In the square lattice we use periodic boundary conditions and in the honeycomb lattice we use use skew boundary conditions in the horizontal
     boundaries.
     }
    \end{center}
\end{figure}

In analogy with simple Ising magnets, in order to measure the paramagnetic-antiferromagnetic phase transition, we shall use the 
{\em staggered magnetisation} as order-parameter
    \begin{equation} \label{4}
    \langle m\rangle =  \frac{1}{N} \left\langle\left|\sum_i c_i\sigma_i\right|\right\rangle,
    \end{equation}
where $N$ is the total number of lattice sites, 
$c_i$ takes values of $\pm 1$ depending in which sub-lattice the site is located (see Fig.~\ref{redes}) and $\langle\ldots\rangle$
stands for time average taken in the stationary regime.

The (staggered) susceptibility is given by
     \begin{equation}
     \chi = N x \{\langle m^2\rangle-\langle m\rangle^2\}.
     \label{susceptibility}
     \end{equation}
We shall use the method proposed in Ref.~\cite{Perez2005}, where three different cumulants are used for the
evaluation of the critical point: (i) the fourth-order or Binder cumulant~\cite{Binder1981}
    \begin{equation}
    U^{(4)}=1-\frac{\langle m^4\rangle}{3\langle m^2\rangle^2},
    \label{cumulant4}
    \end{equation}
(ii) the third-order cumulant (where $\langle m^3\rangle$ is defined analogously to eq.~(\ref{4}))
    \begin{equation}
    U^{(3)}=1-\frac{\langle m^3\rangle}{2\langle m\rangle\langle m^2\rangle},
    \label{cumulant3}
    \end{equation}
and (iii) the second-order cumulant
    \begin{equation}
    U^{(2)}=1-\frac{2\langle m^2\rangle}{\pi\langle m\rangle^2}.
    \label{cumulant2}
    \end{equation}
The scaling forms for the thermodynamic observables, in the stationary state, 
and together with the leading finite-size correction exponent $\omega$ are given by
    \begin{eqnarray}
    m(\epsilon,L) &\approx&  L^{-\beta/\nu}(\hat{M}(\epsilon L^{1/\nu})+
          L^{-\omega} \hat{\hat{M}}(\epsilon L)),\\
    \chi(\epsilon,L) &\approx&  L^{\gamma/\nu}(\hat{\chi}(\epsilon 
          L^{1/\nu})+L^{-\omega} \hat{\hat{\chi}}(\epsilon L)),\\
    U^{(p)}(\epsilon,L) &\approx& \hat{U}^{(p)}(\epsilon L^{1/\nu})+
          L^{-\omega} \hat{\hat{U}}^{[p]}(\epsilon L).\label{scaling3}
    \end{eqnarray}
where $\epsilon=x-x_c$ is the distance from criticality, $p=2$, $3$ or $4$, and $L=\sqrt{N}$ is the linear size of the lattice. 
The parameters $\beta$, $\gamma$ and $\nu$ are the critical exponents for the infinite system, see \cite{Henkel08} for details.  

In principle, the critical point $x_c$ is found from the crossing points in the cumulants $U^{(p)}$. 
A precise estimation of $x_c$ is achieved by taking into account the crossing points for different cumulants $U^{(p)}$ and $U^{(q)}$ with
$p\ne q$ arise for different values of $L$.  
The values of $x$, where the cumulant curves $U^{(p)}(x)$ for two different linear sizes 
$L_i$ and $L_j$ intercept are denoted as $x^{(p)}_{ij}$.
We expand Eq.~(\ref{scaling3}) around $\epsilon=0$ to obtain
    \begin{equation}
    U^{(p)}\approx U^{(p)}_\infty+\bar{U}^{(p)} \epsilon L^{1/\nu} + 
        \bar{\bar{U}}^{(p)} L^{-\omega}+{\rm O}(\epsilon^2,\epsilon L^{-\omega}),
    \label{crossing1}
    \end{equation}
where $U^{(p)}_\infty$ are universal quantities, but $\bar{U}^{(p)}$ and 
$\bar{\bar{U}}^{(p)}$ are non-universal. The value of $\epsilon$ where 
the cumulant curves $U^{(p)}$ for two different linear sizes $L_i$ and 
$L_j$ intercept is denoted as $\epsilon^{(p)}_{i,j}$. At this crossing 
point the following relation must be satisfied:
     \begin{equation}
     L_i^{1/\nu}\epsilon^{(p)}_{ij}+B^{(p)} L_i^{-\omega} =
     L_j^{1/\nu}\epsilon^{(p)}_{ij}+B^{(p)} L_j^{-\omega}.
     \end{equation}
Here $B^{(p)}:=\bar{\bar{U}}^{(p)}/\bar{U}^{(p)}$. Combining for different cumulants ($q\ne p$) we get
    \begin{equation}
    \frac{x^{(p)}_{ij}+x^{(q)}_{ij}}{2}=x_{c}-(x_{ij}^{(p)}-x_{ij}^{(q)})A_{pq},
    \label{linealcrit}
    \end{equation}
where $A_{pq}=(B^{(p)}+B^{(q)})/[2(B^{(p)}-B^{(q)})]$ and is non-universal (see Refs.~\cite{Perez2005,AcunaLara2012} for additional details).
Equation~(\ref{linealcrit}) is a linear equation 
that makes no reference to $\nu$ or $\omega$ and requires as inputs only
the numerically measurable crossing couplings $x_{i,j}^p$. The intercept with the ordinate gives the critical point location.

\section{Results}
\begin{figure}[h!]
    \begin{center}
     \includegraphics[width=6.0cm,clip]{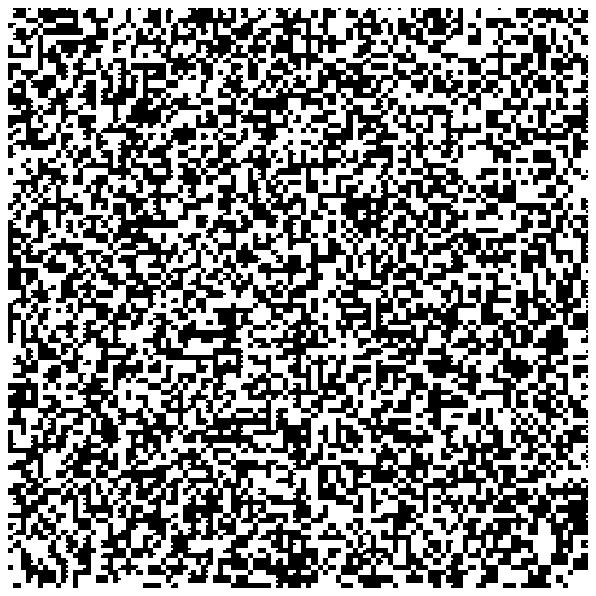}
     \includegraphics[width=6.0cm,clip]{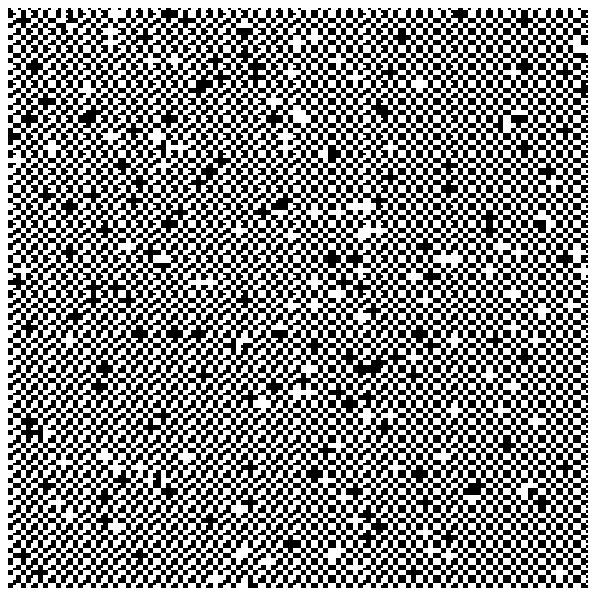}
     \caption[fig1]{\label{snapshots}
     Snapshots of the square lattice with size $L=256$, for a $x=0.1$ (left) and $x=0.95$ (right).
     The order-parameter values are $\langle m\rangle \simeq 0.0034$ and $\langle m\rangle \simeq 0.9446$, respectively.
    }
    \end{center}
    \end{figure}

As a first step, we check that the dynamics Eqs.~(\ref{transition},\ref{3}) 
gives rise to a pa\-ra\-mag\-ne\-tic-antiferromagnetic phase transitions, depending
on the value of the parameter $x>0$. For a qualitative illustration, in
Fig.~\ref{snapshots} we present snapshots for two different values of $x$, of the state of a square lattice of size $L=256$. 
Clearly, the two states look very different and are also distinguished by the respective values of $\langle m\rangle$ as defined in (\ref{4}); 
this suggests the existence of a phase transition, at some intermediate critical point $x_c$. 

We performed simulations on three different lattices with linear sizes
$L=24$, 28, 32, 36, 40 and 48; and did this for both the square and honeycomb lattices, see figure~\ref{redes}.
Starting with a random configuration of spins, the system evolves following the dynamics given by eqs.~(\ref{transition},\ref{3}). 
Then, we let the system evolve during a transient time, that varied from $4\times10^5$ Monte Carlo time steps (MCTS) for $L=24$ to $1.2\times10^6$ 
MCTS for $L=48$. Averages of the observables were taken over $2\times 10^6$ MCTS for $L=24$ and up to $1.2\times 10^7$ MCTS for $L=48$. 
Additionally, for each value of $x$ and $L$, we performed up to 400 independent runs, in order to improve the statistics. 

\begin{figure}[h!]
    \begin{center}
     \includegraphics[width=10.5cm,clip]{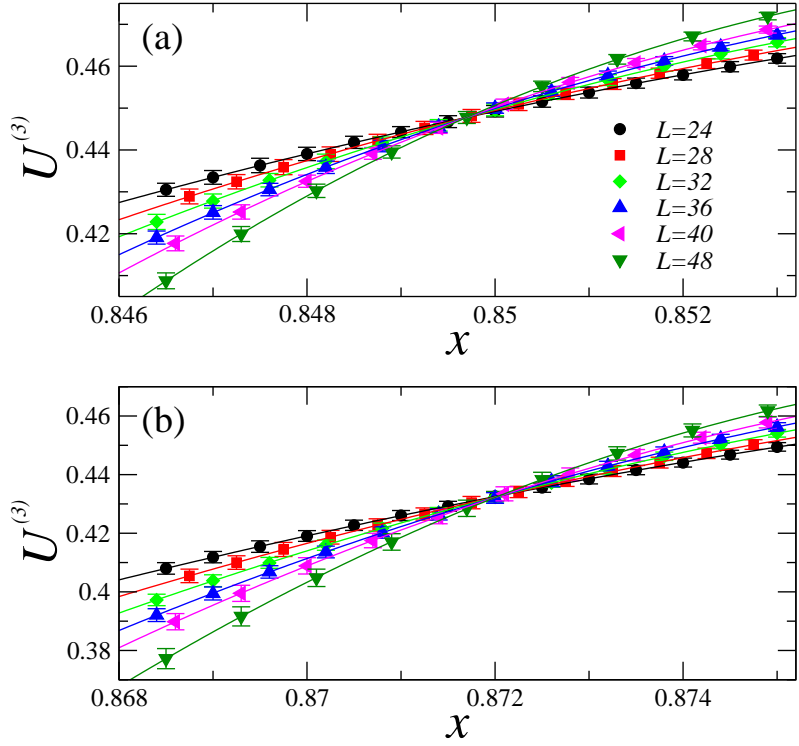}
     \caption[fig2]{\label{cumulantes}
     Third-order cumulant $U^{(3)}$ as function of the parameter $x$ in (a) the square and (b) honeycomb lattices. 
     The solid lines are the third-order polynomial fits.
    }
    \end{center}
    \end{figure}

In Fig.~\ref{cumulantes} we show the third-order cumulant curves as function of $x$ for the square and for the honeycomb lattice. 
The figure shows clearly that in both geometries
the curves for different sizes cross around a certain (lattice-dependent) value of $x$. 
Similar behaviour has been observed as well for $U^{(2)}$ and for $U^{(4)}$.

\begin{figure}[h!]
    \begin{center}
    \includegraphics[width=10.5cm,clip]{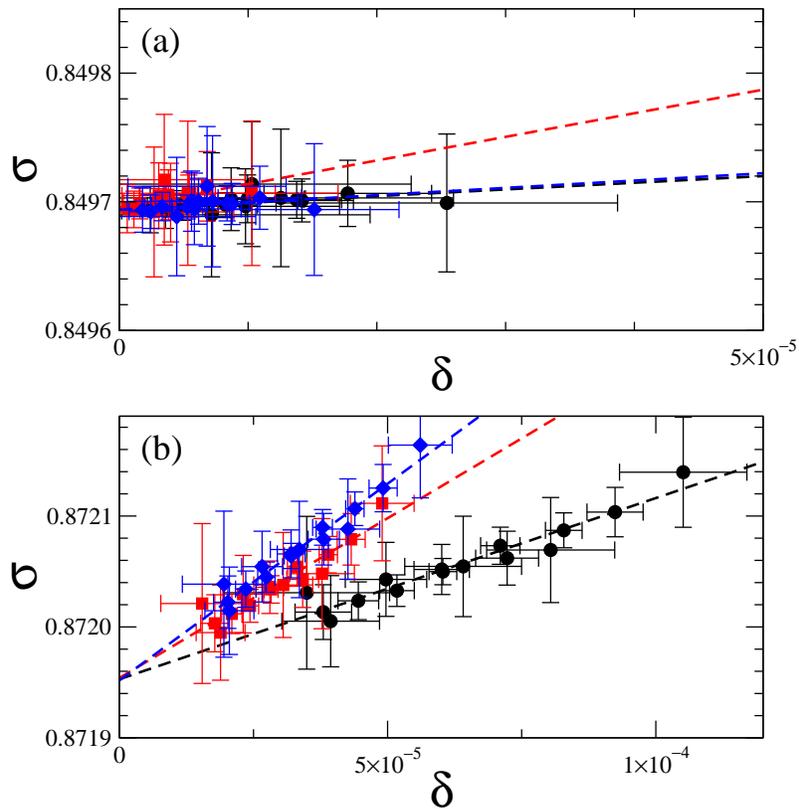}
    \caption[fig3]{\label{criticalpoint} (Color online) Evaluation of the critical 
    point $x_c$ for (a) the square lattice and (b) the honeycomb lattice.
    The points represent the numerical data obtained from third-order polynomial fits and the 
    dashed lines are the linear fits of Eq. (\ref{linealcrit}). Smaller $\delta$-values correspond
    to larger system sizes. 
    Black circles: crossing of $U^{(2)}$-$U^{(4)}$, red squares: crossing of $U^{(2)}$-$U^{(3)}$, 
    blue diamonds: crossing of $U^{(3)}$-$U^{(4)}$.
    }
    \end{center}
    \end{figure}

For the evaluation of the critical points, we used a third-order polynomial fit 
for the cumulant curves. Recalling eq.~(\ref{linealcrit}), the estimation of the
critical points for the two cases are shown in Figure~\ref{criticalpoint}, where we plot
the variable $\sigma :=(x_{ij}^{(4)}+x_{ij}^{(2)})/2$ over against the variable $\delta:=x_{ij}^{(4)}-x_{ij}^{(2)}$. 
We can observe that the raw differences between crosses, $\delta$, are smaller in the square case. But, since 
the smaller values in $\delta$ are around $2\time 10^{-6}$ for the square lattice and $2\time 10^{-5}$ for the honeycomb lattice (these values
correspond to the crossings between the largest sizes using in our simulation), 
we can be sure that largest sizes will not improve significantly our results. In any case, the largest source of numerical error apparently
comes from the statistical uncertainty of the data points. 
If we take another pair of cumulants, the definitions of $\sigma$ and $\delta$ are adapted in an obvious way.  
The linear fits of Eq. (\ref{linealcrit}) give the following estimates for the critical points
\begin{equation}
\left\{ \begin{array}{ll} x_c=0.84969(4)  & \mbox{\rm ~~;~ square lattice} \\
                          x_c=0.87195(14) & \mbox{\rm ~~;~ honeycomb lattice}
        \end{array} \right.
\end{equation}
where the numbers in brackets give the estimated uncertainty in the last given digit(s). 
Both results are in good agreement with the reported values for the ferromagnetic MVM~\cite{Oliveira1992,Kwak2007,AcunaLara2014}.
We want to point out that, when we compare the data for the crossing of the $U^{(2)}-U^{(4)}$ 
curves with the previous reported data for the ferromagnetic case
from Ref.~\cite{AcunaLara2014}, the range in the differences $\delta$ is almost two times larger in the ferromagnetic case. 
Since the linear sizes considered are the same
in both cases, we suspect that the scaling effects are smaller in the antiferromagnetic case. 
Additional simulations in the Ising model, for antiferromagnetic and
ferromagnetic, interactions for different lattice geometries and for the MVM on square lattices would help to see if this is a universal behaviour.

\begin{figure}[h!]
    \begin{center}
    \includegraphics[width=10.5cm,clip]{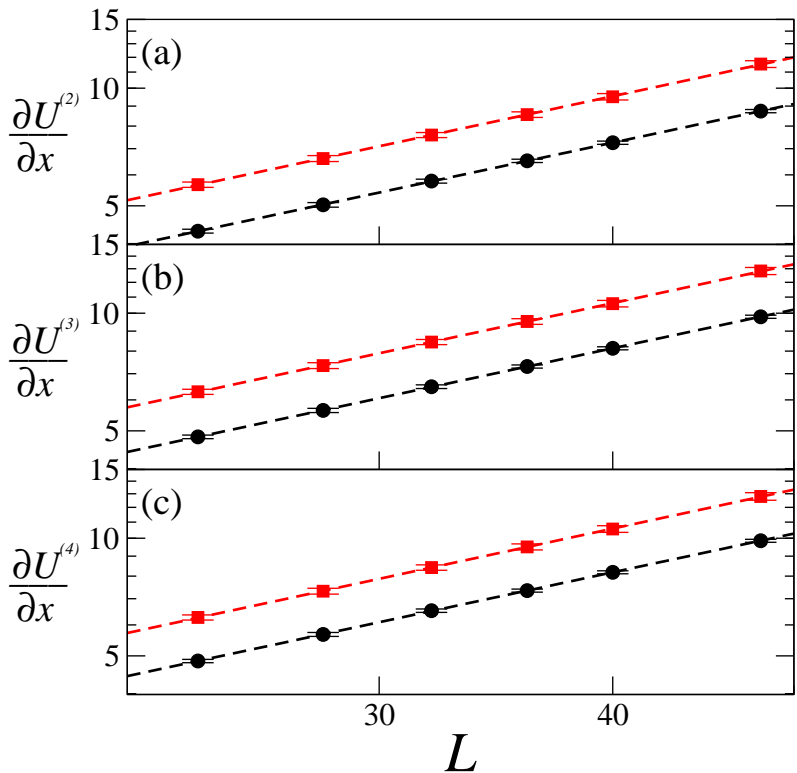}
    \caption[fig4]{\label{nuexponent} (Color online)
    Log-log plot of the derivatives of the cumulant (a) $\partial U^{(2)}/\partial x$, (b) $\partial U^{(3)}/partial x$ 
    and (c) $\partial U^{(4)}/partial x$, taken {\em at} the critical point $x=x_c$. 
    Black circles correspond to the square lattice and red squares to the honeycomb lattice. 
    The dashed lines are power-law fits.
    }
    \end{center}
    \end{figure}

\begin{figure}[h!]
    \begin{center}
    \includegraphics[width=10.5cm,clip]{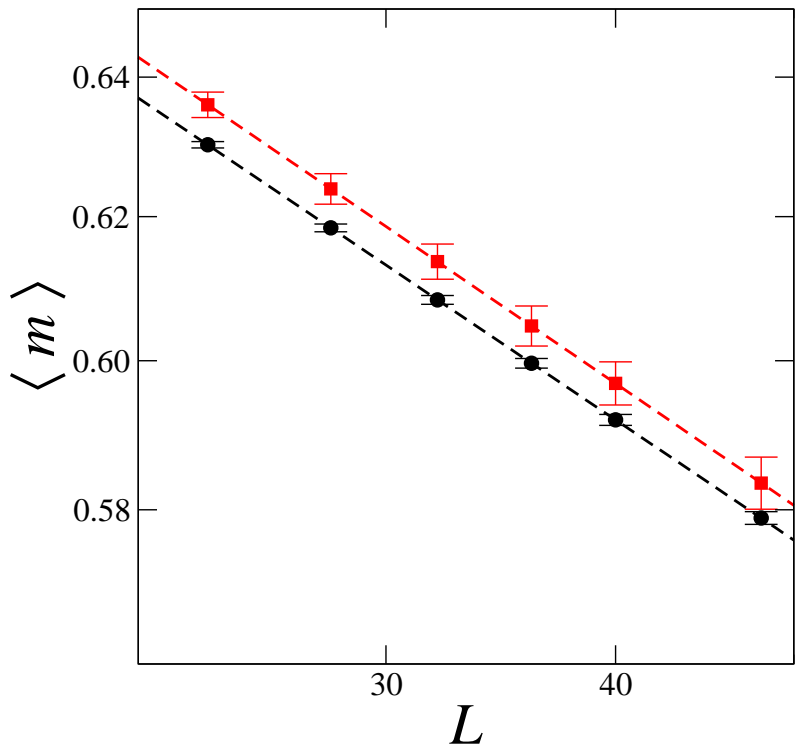}
    \caption[fig6]{\label{betaexponent} (Color online)
    Log-log plot of the order parameter at the critical point
    for the square (black circles) and honeycomb (red squares) lattices.
    The dashed lines are power-law fits.
    }
    \end{center}
    \end{figure}

The critical exponents can be evaluated, by using Eqs.~(\ref{scaling3}), at the critical point $\epsilon=0$. 
One expects the following finite-size scaling behaviour 
    \begin{eqnarray}
    m(L) &\propto& L^{-\beta/\nu}, \label{beta}\\
    \chi(L) &\propto& L^{\gamma/\nu}, \label{gamma}
    \end{eqnarray}
and
    \begin{equation}
    \frac{\partial U^{(p)}}{\partial x}\Bigl|_{x=x_c} 
        \propto L^{1/\nu},
    \label{nu}
    \end{equation}
In Fig.~\ref{nuexponent}, we show the derivatives of the cumulants at the critical point. From the finite-size scaling law~(\ref{nu}),
we obtain the following results: $1/\nu=1.02(4)$ for 
the three cumulants in the square lattice 
and $1/\nu=1.02(4)$, 1.03(4) and 1.03(4) for $U^{(2)}$, $U^{(3)}$ and $U^{(4)}$ respectively in the honeycomb case.
The evaluation of $\gamma/\nu$ is shown in Fig.~\ref{gammaexponent}, with the relation (\ref{gamma}) 
we obtain $\gamma/\nu=1.756(8)$ and $\gamma/\nu=1.759(10)$ for the square and honeycomb lattices respectively. 
We present in Fig.~\ref{betaexponent} the evaluation of $\beta/\nu$, with Eq.~(\ref{beta}) we obtain $\beta/\nu=0.123(2)$ 
and $\beta/\nu=0.124(8)$ for the square and honeycomb lattices, respectively. 

    \begin{figure}[h!]
    \begin{center}
     \includegraphics[width=10.5cm,clip]{fig_05.eps}
     \caption[fig5]{\label{gammaexponent} (Color online)
     Log-log plot of the susceptibility at the critical point
     for the square (black circles) and honeycomb (red squares) lattices.
     The dashed line are power-law fits.}
    \end{center}
    \end{figure}

Our results are collected in table~\ref{totales}, where we also included the corresponding results for the ferromagnetic MVM \cite{Kwak2007,AcunaLara2014}. 
Clearly, all reported numerical estimates are very close to the exactly known values of the two-dimensional Ising ferromagnet, 
e.g. \cite[appendix A]{Henkel08}. 
According to hyperscaling, one expects $(2\beta+\gamma)/\nu=d=2$, which is well confirmed by the simulation results. 
We also see that the agreement for the anti-ferromagnetic MVM is even slightly better than for the ferromagnetic MVM. 
It turned out that an explicit consideration of possible finite-size corrections, described by the Wegner exponent $\omega$, was not necessary. 
\begin{center}
\begin{table}[ht]
\caption[tab1]{\label{totales} Values of the critical parameters for the antiferromagnetic Majority voter model (AMVM). 
For comparison, we also include the values reported for the ferromagnetic Majority voter model (FMVM) 
by Kwak~{\it et al.}~\cite{Kwak2007} on square lattices and by Acu\~na-Lara~{\it et al.}~\cite{AcunaLara2014}
on honeycomb lattices. 
}
\begin{tabular}{@{}llllll@{}} \hline \hline
\mcc{$x_c$} & \mcc{$1/\nu$} & \mcc{$\gamma/\nu$} & \mcc{$\beta/\nu$} & \mcc{$(\gamma+2\beta)/\nu$}        & ~~model \\ \hline
0.84969(4)  & 1.02(4)       & 1.756(8)           & 0.123(2)          & ~2.002(9)                          & Square AMVM \\
0.87195(14) & 1.03(4)       & 1.759(10)          & 0.124(8)          & ~2.007(15)                         & Honeycomb AMVM \\
0.8500(4)   & 0.98(3)       & 1.78(5)            & 0.120(5)          & ~2.02(5)                           & Square FMVM \\
0.8721(1)   & 1.01(2)       & 1.755(8)           & 0.123(2)          & ~2.001(9)                          & Honeycomb FMVM \\
\hline\hline
\end{tabular}
\end{table}
\end{center}

\begin{figure}[h!]
    \begin{center}
    \includegraphics[width=10.5cm,clip]{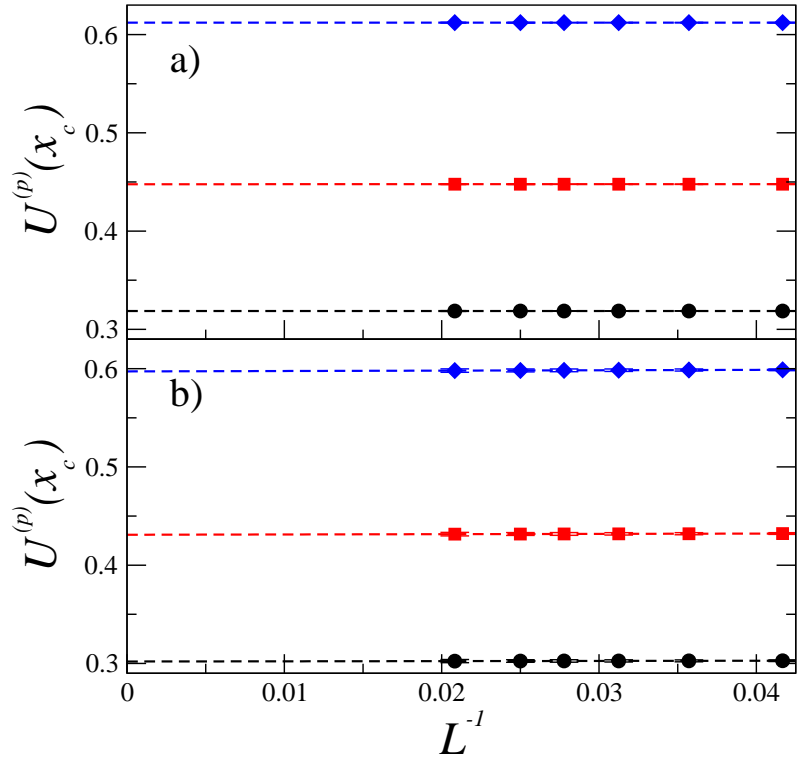}
    \caption[fig6]{\label{bindergraf} (Color online)
    Cumulant values at the critical point as function of the inverse linear size for a) the
    square and b) honeycomb lattices. Black circles: $U^{(2)}$, red squares: $U^{(3)}$, blued diamonds : $U^{(4)}$
    The dashed lines are linear fits.
    }
    \end{center}
    \end{figure}

Additionally, we evaluated the universal quantities (lattice-dependent) $U^{(p)}_\infty$ for both lattices, using Eq.~(\ref{scaling3})
with $\epsilon=0$. Our data do not allow to reliably evaluate the Wegner exponent $\omega$, 
since there is no need to include scaling corrections in these
systems, as we illustrate in Fig.~\ref{bindergraf}.
Our result for the Binder cumulant, $U^{(4)}_\infty=0.612(1)$, for the square lattice is in good agreement 
with the reported values of the two-dimensional Ising model,
$U^{(4)}=0.610690(2)$ \cite{Kamieniarz1993}, or $U^{(4)}=0.6106922(16)$  \cite{Salas2000}. 
We listed all the results for the three cumulants, for both lattices,  in Table~\ref{bindertabla}.
We observe that the results for the honeycomb and square lattices are clearly different from each other, in all cases. 
The explanation for this behaviour is simple: because in our
simulations, the boundary conditions of the two lattices are different: 
we use periodic boundary conditions for the square lattice and skew conditions in the horizontal boundaries for the
honeycomb lattice (see Fig.~\ref{redes}). Estimates of $U^{(4)}$ reported in the literature, and based on simulations 
in two-dimensional critical Ising model, produced a similar spread  
for different lattices or boundary conditions \cite{Selke2006}. 
%
In order to test the universality through the cumulants $U^{(p)}$, 
it would be necessary to perform simulations for the Ising model on the honeycomb lattice with the same
boundary conditions.
\begin{center}
\begin{table}
\caption[tab2]{\label{bindertabla} Cumulants for the square and honeycomb lattices obtained in this work. 
}
\center{
\begin{tabular}{llll} \hline \hline
\mcc{$U^{(2)}_\infty$} & \mcc{$U^{(3)}_\infty$} & \mcc{$U^{(4)}_\infty$}  & ~Geometry \\ \hline
        0.319(1)       &        0.448(1)        &          0.612(1)       & Square    \\
        0.302(2)       &        0.431(2)        &          0.597(2)       & Honeycomb \\
\hline\hline
\end{tabular}
}
\end{table}
\end{center}

\section{Conclusions}
We have introduced an antiferromagnetic version of the MVM, both on the two-dimensional square and honeycomb lattices, and studied
its stationary state, as a function of the control parameter $x$, through intensive Monte Carlo simulations. 
The phase transition in the stationary state, between an ordered phase for $x$ large enough and a disordered phase for $x$ small enough, 
falls on both lattices into the $2D$ Ising model universality class, and independently of whether the number of nearest neighbours is even or odd.  
This model further illustrates that the set of critical exponents is consistent in two-dimensional regular lattices. 
Future work should explore if additional phases and/or multicritical points exist if an external field is included.

\section*{References}

\bibliography{bibliografia}

\end{document}